\documentclass[pra,10pt,superscriptaddress,aps,twocolumn,longbibliography,nofootinbib,floatfix]{revtex4-2}
\pdfoutput=1
\usepackage{graphicx}
\usepackage{xcolor}
\usepackage{amsmath}
\usepackage{amssymb}
\usepackage{amsthm}
\usepackage{mathtools}

\usepackage{enumitem}
\usepackage[T1]{fontenc}
\usepackage{bbm}
\usepackage{dsfont}
\usepackage[linesnumbered,ruled,vlined]{algorithm2e}
\SetKwInput{kwInit}{Init}
\usepackage{tikz}
\usetikzlibrary{positioning}
\usepackage[caption=false]{subfig}
\usepackage{siunitx}
\usepackage{qcircuit}
\usepackage{pifont}
\usepackage[colorlinks=true, citecolor=blue, linkcolor=red]{hyperref}


\newcommand{\Tr}{\text{Tr}}

\renewcommand{\eqref}[1]{(\ref{#1})}
\newcommand{\rhobar}{\bar{\rho}}

\newtheoremstyle{example}{\topsep}{\topsep}%
{}%         Body font
{}%         Indent amount (empty = no indent, \parindent = para indent)
{\bfseries}% Thm head font
{:}%        Punctuation after thm head
{   }%     Space after thm head (\newline = linebreak)
{\thmname{#1}\thmnumber{ #2}}%\thmnote{ #3}}%         Thm head spec
\theoremstyle{example}
\theoremstyle{definition}

\newtheorem*{theorem*}{Theorem}

\def\orcid#1{\kern -0.4em\href{https://orcid.org/#1}{\includegraphics[keepaspectratio,width=0.7em]{orcid_logo.pdf}}}

\renewcommand{\>}{\rangle}
\newcommand{\<}{\langle}

\long\def\ca#1\cb{}

\begin{document}
\title{Opportunities and challenges in scaling quantum error detection on hardware}

\author{Yanis Le Fur}
% \thanks{Corresponding author: \href{mailto:yanis.lefur@epfl.ch}{yanis.lefur@epfl.ch}.}
\affiliation{Institute of Physics, École Polytechnique Fédérale de Lausanne (EPFL), Lausanne, Switzerland}
\affiliation{Departamento de Física Téorica de la Materia Condensada and Condensed Matter Physics Center (IFIMAC),
Universidad Autónoma de Madrid, 28049 Madrid, Spain}
\affiliation{Institute of Fundamental Physics IFF-CSIC, Calle Serrano 113b, 28006, Madrid, Spain}
\author{Ethan Egger}
\affiliation{Department of Computer Science and Engineering, Michigan State University, East Lansing, MI 48823, USA}
\affiliation{Department of Physics and Astronomy, Michigan State University, East Lansing, MI 48823, USA}
\affiliation{Center for Quantum Computing, Science, and Engineering, Michigan State University, East Lansing, MI 48823, USA}
\affiliation{Center for Quantum Information and Control, University of New Mexico, Albuquerque, NM 87131, USA}

\author{Hong-Ye Hu}
\affiliation{Department of Physics, Harvard University, Cambridge, MA 02138, USA}
\affiliation{EdenCode Inc.}

\author{Vincent Russo}
\affiliation{Unitary Foundation}

\author{William J. Zeng}
\affiliation{Unitary Foundation}
\affiliation{Quantonation}

\author{Ryan LaRose}
\thanks{Corresponding author:  \href{rmlarose@msu.edu}{rmlarose@msu.edu}}
\affiliation{Department of Computational Mathematics, Science, and Engineering, Michigan State University, East Lansing, MI 48823, USA}
\affiliation{Department of Electrical and Computer Engineering, Michigan State University, East Lansing, MI 48823, USA}
\affiliation{Department of Physics and Astronomy, Michigan State University, East Lansing, MI 48823, USA}
\affiliation{Center for Quantum Computing, Science, and Engineering, Michigan State University, East Lansing, MI 48823, USA}

\begin{abstract}
    Quantum error detection can produce unbiased expectation values that exponentially converge to noiseless results as the code distance is increased. Despite this, its performance as an error mitigation technique is relatively understudied on quantum hardware because of its two main drawbacks: (i) the number of samples increases exponentially in the circuit depth/noise level, and (ii) the classical processing generally grows exponentially in the code distance, though exceptions exist. Additionally, the constant (but often large) overhead of embedding the code and logical operations on hardware can make accuracy worse instead of better. In this work, we seek to provide a clear picture of these opportunities and challenges for scaling quantum error detection on hardware. We do so by performing a detailed benchmarking study on real and simulated noisy quantum computers, using the repetition code and triangular color code for memory experiments and logical computations with up to $74$ physical qubits. In addition to these benchmarks, we estimate the pseudothreshold of codes to map the frontier of error detection on current and future quantum computers. Despite the challenges, our results show strong promise for scaling quantum error detection on hardware.
\end{abstract}

\maketitle

% \tableofcontents

% =============================================================================
\section{Introduction}
\label{sec:introduction}
% =============================================================================

Due to the experimental fragility of quantum states, it is generally accepted
that quantum computers must employ active and/or passive methods to deal with
errors. Recent landmark experiments in quantum error correction have
demonstrated logical error rates that decrease exponentially in the number of
physical qubits used~\cite{google2021exponential, google2023suppressing, google2025quantum}. These
error rates are primarily in ``memory experiments'' in which a logical state is
prepared and then stored on the device for some time. In recent
years, there has also been an explosion in the development of so-called quantum error
mitigation methods~\cite{cai2023quantum,russo2026quantum}.
These error mitigation methods are generally designed to improve the accuracy or
precision of computing expectation values on (current) quantum computers. While some error mitigation methods use additional space resources (i.e., qubits) --- for example virtual distillation~\cite{huggins2021virtual} and recent proposed improvements to probabilistic error cancellation~\cite{DalFavero_LaRose_2025} --- most use additional time, guided by the limitations of current quantum computers. However, recent experimental findings in quantum error correction demonstrate that combining multiple noisy qubits can improve overall system performance. This motivates the potential value of using additional spatial resources for quantum error mitigation.

Indeed, the error mitigation technique of quantum error detection shares many desirable features with quantum error correction. In quantum error detection, additional qubits are used to encode logical states, which are then evolved by logical operations. However, instead of attempting to actively correct errors, the logical information is measured and then post-processed to improve expectation values as in error mitigation.  Intuitively, the method
works by discarding final states that are not in the logical subspace of the chosen error correcting code. Error detection and variants have been proposed under the names of subspace
expansion~\cite{mcclean2017hybrid, mcclean2020decoding, huggins2021efficient,
yoshioka2022generalized}, error detection~\cite{knill2005quantum,
corcoles2015demonstration, linke2017fault}, symmetry
verification~\cite{mcardle2019error, bonet2018low, cai2021quantum}, and logical
shadow tomography~\cite{hu2022logical}. Experimentally, the method has been used in~\cite{corcoles2015demonstration, urbanek2020error, gong2022experimental, self2024protecting, chertkov2025error, martiel2025low, javadi2025big, vezvaee2026demonstration, zhang2026quantum}. The main challenges are that the method requires a number of samples (shots) exponential in the depth of the quantum circuit under most reasonable noise models, and that the classical pre/post-processing generally scales exponentially in the code size (a problem which several authors have attempted to alleviate~\cite{yoshioka2022generalized, hu2022logical}), though exceptions (e.g. CSS codes) exist.

Despite these challenges, quantum error detection has very desirable characteristics. Most notably, while virtually all error mitigation techniques are heuristic and produce biased estimators,  quantum error detection produces unbiased estimates of expectation values, assuming all errors which occur are correctable. To our knowledge, the only other unbiased
quantum error mitigation technique is probabilistic error cancellation (PEC)~\cite{temme2017error, endo2018practical, zhang2020error, van2023probabilistic}, assuming noise can be perfectly characterized. In spite of this, quantum error detection has been relatively understudied on hardware in general, and in particular with respect to its performance as the code size scales. 

In this work, we seek to fill this gap by presenting a detailed benchmarking study of quantum error detection on real and simulated quantum computers, highlighting both the opportunities and challenges of the method. Our results demonstrate that the accuracy of expectation values can (exponentially) converge as the code distance/number of physical qubits increases, analogous to recent experiments in quantum error correction. To demonstrate this, we study the performance of the repetition code and the triangular color code on IBM quantum computers, analyzing the accuracy of expectation values as the code size increases. Our experiments use up to $74$ physical qubits and $3146$ physical two-qubit gates to prepare logical Bell states, providing better-than-physical performance when evaluating expectation values. We also provide a clear picture of the sampling overhead required and the amount of classical pre/post processing required to achieve these results, and characterize the pseudothreshold (i.e., when error detection provides an advantage over the physical experiment) for preparing a logical Bell state with the triangular color. Although sampling and classical processing remain challenges for the method that we do not seek to address in this work, our experimental results indicate the opportunities for error detection at the scale of Megaquop computers~\cite{Preskill_2025} and beyond. To these ends, we provide background on quantum error detection in Sec.~\ref{sec:preliminaries}, we present our experimental results in Sec.~\ref{sec:results}, and we discuss our methods in Sec.~\ref{sec:methods}.

% =============================================================================
\section{Preliminaries}
\label{sec:preliminaries}
% =============================================================================

% =============================================================================
\subsection{Quantum error detection}
\label{sec:quantum-error-detection}
% =============================================================================

Quantum error detection encodes a physical state into a logical state using an error correction code, implements logical operations, then projects the state into the codespace. 
Specifically, given a physical state $\rho$, an observable $O$, and an $[[n, k, d]]$ error correction code defined by stabilizer code $\mathcal{S} = \langle S_1, ..., S_r \rangle$, the error-mitigated expectation value is given by
\begin{equation} \label{eqn:error-mitigated-value}
    \langle O \rangle := \frac{\Tr \left [ \Pi \bar{\rho} \Pi^\dagger \bar{O} \right]}{ \Tr \left[ \Pi \bar{\rho} \Pi^\dagger \right]}
\end{equation}
where the projection onto the logical codespace $\Pi$ is defined from the code via
\begin{equation} \label{eqn:projection-onto-codespace-pi}
    \Pi := \prod_{i = 1}^{r} (I + S_i) / 2 = \frac{1}{2^r} \sum_{S \in \mathcal{S}} S.
\end{equation}
Here, $n$ is the number of physical qubits encoding $k$ logical qubits, the number of stabilizer generators is $r = n - k$, and $d$ is the distance of the code. This projector $\Pi$ has the following properties to mitigate errors. Given a correctable error $E$ acting on a logical state $|\bar{\psi}\rangle$, there exists some stabilizer generator $S$ such that $ES = - SE$. Therefore we have $\frac{I + S}{2} E |\bar{\psi}
\rangle = 0$.
%
% \begin{align*}
%     \frac{I + S}{2} E |\bar{\psi} \rangle &\propto E |\bar{\psi} \rangle + SE |\bar{\psi} \rangle \\
%         &= E |\bar{\psi} \rangle - E S |\bar{\psi} \rangle \\
%         &= E |\bar{\psi} \rangle - E |\bar{\psi} \rangle \qquad (\text{since } S |\bar{\psi} \rangle = |\bar{\psi} \rangle \text{by def. of stabilizer}) \\
%         &= 0.
% \end{align*}
%
On the other hand, if there is no error, then $\frac{I + S}{2} |\bar{\psi}
\rangle = |\bar{\psi}\rangle$. Thus, the operator $\Pi$ projects or filters out
correctable errors, while keeping codewords. 

From these properties, it follows that the error-mitigated expectation
value~\eqref{eqn:error-mitigated-value} is unbiased if all errors that occur are
correctable. While in practice, bias enters due to uncorrectable errors,
including logical errors, this is still a highly desirable property for a
quantum error mitigation technique. As mentioned, the only other unbiased
quantum error mitigation technique to our knowledge is probabilistic error cancellation (PEC)~\cite{temme2017error, endo2018practical, zhang2020error, van2023probabilistic}. Like
error detection, this unbiased property in probabilistic error cancellation also
depends on an assumption about the noise --- namely, that it can be characterized to
arbitrary accuracy.

Under most noise models, the probability of measuring codewords becomes exponentially small in the circuit depth. For example, consider a global depolarizing channel $\rhobar \mapsto p \rhobar + (1 - p) I / 2^n$ after each layer of the circuit. Here, $p$ is the operation layer fidelity, $0 \le p \le 1$, and $I / 2^n$ is the maximally mixed state of the same dimension as $\rhobar$ ($2^n)$. After $D$ layers, the final state is
\begin{equation} \label{eqn:final-state-D-layer-depolarizing}
    % \rho \mapsto (1 - p)^D \rho + \left[ (1 - p)^{D - 1} + (1-p)^{D - 2} + \cdots + 1 \right] p I / d
    \rhobar \mapsto p^D \rhobar + (1 - p^D) I / 2^n .
\end{equation}
Thus, we see the probability of measuring a codeword scales as $p^D$ ($0 \le p \le 1$).  {This reveals a {main challenge with quantum error detection}: the probability of sampling codewords is exponentially small in the circuit depth $D$ under most noise models. (Equivalently, the number of samples required grows exponentially in the circuit depth.)}
Under common assumptions, all quantum error mitigation techniques have similar characteristics, either requiring exponentially many samples or requiring some other resource which grows exponentially with the problem size~\cite{Takagi_Endo_Minagawa_Gu_2022}. For example, the error mitigation technique of PEC mentioned above requires a number of circuits (samples) that grows exponentially in the circuit depth. This can be understood simply by definition of the protocol --- each unitary in a depth $D$ circuit is expanded as a sum over $m$ basis elements assumed to be implementable on a noisy quantum computer, so the product of all unitaries gets expressed as a summation with $m^D$ terms.

In quantum error detection and correction, a larger code distance $d$ is desirable. This is because a larger distance separates codewords more, making it less likely for logical errors to occur. This means that, in principle, using \textit{more} noisy qubits in quantum error detection \textit{reduces} errors in expectation value estimates. In this work, we present results of multiple benchmarks performed on IBM quantum computers which show this is true: {The {main advantage with quantum error detection} is that accuracy can be improved by using additional physical qubits (larger code distances).}

\begin{figure}
    \centering
    \includegraphics[width=\linewidth]{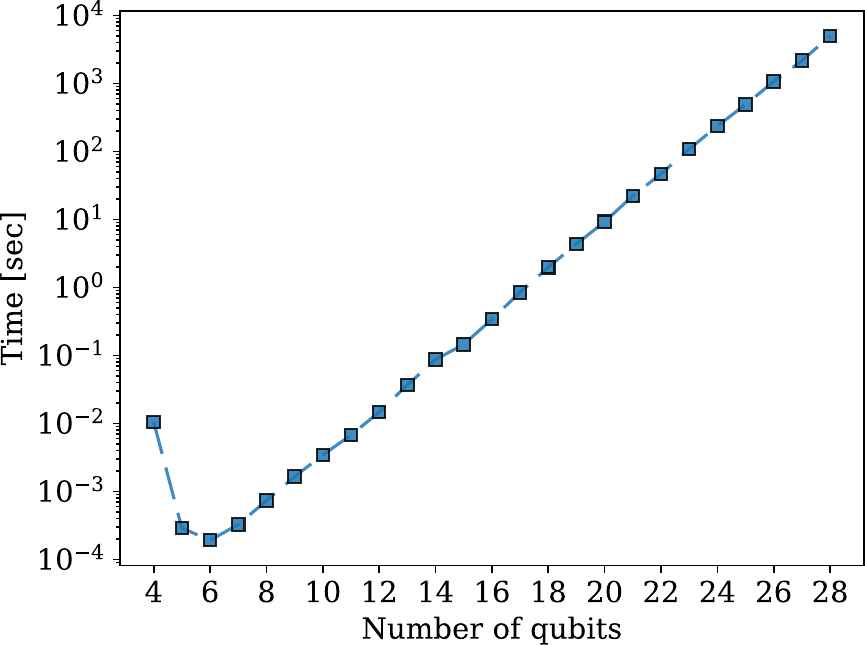}
    \caption{Wall clock time in seconds to compute codewords from stabilizer generators for random codes on $4 \le n \le 28$ qubits. Using the best algorithm to our knowledge (generating a random state vector and multiplying it by $S + I$ for each generator $S$) and the best implementation of this algorithm (in Stim~\cite{Gidney_2021}), we benchmarked the time to compute codewords from stabilizer generators for random codes with $4 \le n \le 28$ physical qubits. The results confirm the exponential scaling and give a sense for the practicality of this strategy: 
    for $n\geq25$ the runtime reaches hours of computation or higher, and beyond $n=28$ the memory requirements exceeded our available system capacities. These complications demonstrate the hard pre-processing required for error detection.}
    \label{fig:stim-codewords-timing}
\end{figure}

However, the classical (post-)processing required by quantum error detection is computationally hard. There are two broad techniques to implement error detection (evaluate~\eqref{eqn:error-mitigated-value}):
\begin{enumerate}
    \item Measure all qubits (in the computational basis) and only keep codewords. This is efficient \textit{assuming the codewords are known}. While computing codewords from stabilizer generators is efficient for certain codes (CSS codes), in general this is a hard problem.  In Fig.~\ref{fig:stim-codewords-timing}, we show runtime of the best-known algorithm to compute codewords from stabilizers. The punchline is that computing codewords from stabilizers scales exponentially and becomes infeasible for relatively small $n \simeq 28$ codes. %, \textit{and therefore quantum error detection using this post-processing technique scales exponentially and quickly becomes infeasible.}

    \item Evaluate the numerator and denominator
    of~\eqref{eqn:error-mitigated-value} as expectation value problems for the
    modified observables $\Pi^\dagger \bar{O} \Pi$ (numerator) and $\Pi$
    (denominator). (Note that as long as $\bar{O}$ and $\Pi$ commute then the modified observable
    for the numerator can be simplified to $\Pi \bar{O}$.)
    This means we can
    implement quantum error detection by estimating the expectation values $\Tr
    [ \rho \Pi \bar{O} ]$ (numerator\footnote{Here we assume a subspace code in which logical observables commute with all stabilizers, which is not true in general for subsystem codes.}) and $\Tr [
    \rho \Pi ]$ (denominator), where the projector onto the codespace $\Pi$ is
    defined in~\eqref{eqn:projection-onto-codespace-pi}.
    Unfortunately, $\Pi$ is not Pauli and so does not admit an efficient stabilizer/tableau
    description. Indeed, $\Pi$ is in general a summation of $2^r$ Paulis (recall that $r = n - k$).
    This means that the time for classical post-processing generally scales as $2^r$.
\end{enumerate}

This reveals {a second main challenge with quantum error detection: the classical (post-)processing scales exponentially in the code size.

\subsection{Literature review of\\quantum error detection experiments}

\begin{table*}
    \scriptsize
    \centering
    \begin{tabular}{c|c|c|c|c|c}
        Year & Problem & Code & Physical qubits & Computer & Ref \\ \hline
        2015 & Bell state preparation & $[[2, 0, 2]]$ & $4$ & Four superconducting qubits & \cite{qed2015} \\
        2020 & Two-qubit Hydrogen ground state & $[[4, 2, 2]]$ & $4 + 2$ ancilla & IBM Tokyo & \cite{Urbanek_Nachman_Jong_2020} \\
        2022 & State preparation & $[[5, 1, 3]]$ & $5$ & 12 superconducting qubits & \cite{Gong2022} \\
        2024 & Mirror circuits/Quantum volume & Iceberg $[[k + 2, k, 2]]$ & $10 + 2$ ancilla & Quantinuum H2 & \cite{Self_Benedetti_Amaro_2024} \\
        2025 & Dissipative quantum dynamics & $[[4, 2, 2]]$ & $28$ & Quantinuum H2 & \cite{Chertkov_Potter_Hayes_2025} \\
        2025 & (Random) State preparation & Spacetime & $50 + 18$ ancilla & IBM Kingston & \cite{Martiel_JavadiAbhari_2025} \\
        2025 & GHZ state preparation & GHZ state & $120 + 8$ ancilla & IBM Aachen (and others) & \cite{JavadiAbhari_Martiel_Seif_Takita_Wei_2025} \\
        2025 & Bell state preparation & $[[4, 2, 2]]$ & $4$& IBM Kyiv & \cite{Vezvaee_Tripathi_MorfordOberst_Butt_Kasatkin_Lidar_2025} \\
        2026 & Bell state preparation & $[[2, 0, 2]]$ & $4$ & Silicon quantum computer & \cite{Zhang_2026} \\
    \end{tabular}
    \caption{Quantum error detection experiments on quantum computers.}
    \label{tab:error-detection-experiments}
\end{table*}

Table~\ref{tab:error-detection-experiments} and the papers we highlight in this section provide a current picture of the experimental status of quantum error detection. The theory of quantum error detection is well established. In an influential 2020 paper, McClean \textit{et al.}~\cite{McClean_Jiang_Rubin_Babbush_Neven_2019} proposed it as an ``intermediate form of error correction'' to enable practical applications, and generalized the projection~\eqref{eqn:projection-onto-codespace-pi} to a subset of symmetries with general coefficients (i.e., modified~\eqref{eqn:projection-onto-codespace-pi} to $\Pi = \sum_i c_i S_i$ for general coefficients $c_i$) that, when measured, leads to a generalized eigenvalue problem. Numerical simulations show improved accuracy for a benchmark with the $[[5, 1, 3]]$ code and for a four-qubit Hydrogen molecule with problem symmetries under a single-qubit depolarizing noise model. Further, recognizing the challenging classical post-processing, the authors propose a stochastic sampling scheme to evaluate the error-mitigated expectation value~\eqref{eqn:error-mitigated-value}. Last, the paper also proposes modifying the projectors in~\eqref{eqn:projection-onto-codespace-pi} as $ (I + (-1)^{s_i} S_i) / 2$ to utilize the $i$th syndrome bit $s_i$ for ``corrections with recovery operations.''

Around the same time, a 2021 paper by Cai explored similar ideas under the names of symmetry verification and symmetry expansion~\cite{Cai_2021}, with the primary difference being that the symmetries in this work are problem symmetries (e.g., symmetries of the Hamiltonian for an energy estimation problem), whereas the symmetries in the previous work (McClean \textit{et al.}~\cite{McClean_Jiang_Rubin_Babbush_Neven_2019}) are either problem symmetries or stabilizers from error correction codes. Cai numerically investigates their performance with up to $n = 12$ qubit 2D Fermi-Hubbard models with depolarizing noise, considering different sets of symmetries arising from the Hubbard model, and showing improved performance.

Several works have performed error detection with the use of ancilla qubits. For example, in 2024 Self \textit{et al.} used ten data qubits and two ancilla qubits to measure stabilizers of the $[[k, k + 2, 2]]$ ``Iceberg code'' on the Quantinuum H2 computer~\cite{Self_Benedetti_Amaro_2024}, demonstrating better-than-physical performance on mirror circuit and quantum volume benchmarks. In 2025, Martiel and Javadi-Abhari used up to $18$ ancilla qubits and up to $50$ data qubits on IBM quantum computers to prepare random stabilizer states, showing a 236x fidelity gain relative to the physical experiment~\cite{Martiel_JavadiAbhari_2025}. This paper is also notable for using codes tailored to hardware and introduce techniques to identify and measure low-weight stabilizers. Additionally in 2025, Javadi-Abhari \textit{et al.} used $8$ ancilla qubits and up to $120$ data qubits to prepare GHZ states~\cite{JavadiAbhari_Martiel_Seif_Takita_Wei_2025}. Last, in 2025 Chertkov \textit{et al.}~\cite{Chertkov_Potter_Hayes_2025} implemented quantum error detection in an experimental study of a dissipative quantum process using the $[[4, 2, 2]]$ code and two ancilla to measure stabilizers.

In 2025, Chertkov \textit{et al.}~\cite{Chertkov_Potter_Hayes_2025} implemented quantum error detection in an experimental study of a dissipative quantum process. This process contains mid-circuit resets that are applied at random locations/times. In their work, the authors perform a logical version of the computation with each pair of physical qubits encoded with the $[[4, 2, 2]]$ code. Using two ancilla qubits to measure the stabilizer generators of this code, they implement a reset whenever an error is detected via ancilla measurement. In this sense, the authors ``avoid the exponential post-selection [sampling] penalty of quantum error detection by using an adaptive quantum circuit whose dynamics are controlled by the mid-circuit error detection
measurements''. In other words, the authors of this paper similarly identify the exponential sampling overhead of quantum error detection. While they address it, they do so for a particular quantum circuit and application, and not for error detection in general.

% =============================================================================
\section{Results}
\label{sec:results}
% =============================================================================

\begin{figure*}
    \centering
    \includegraphics[width=\linewidth]{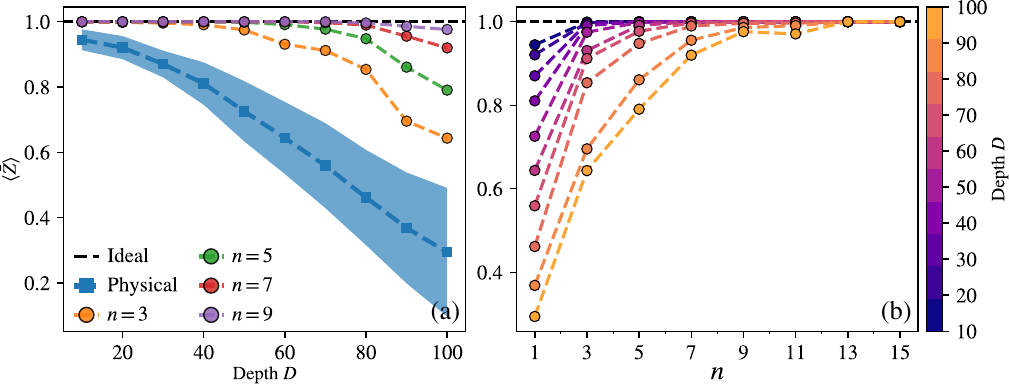}
    \caption{Repetition code memory experiment results on IBM Kyiv. Subplot (a) shows $\langle \bar{Z} \rangle$ versus circuit depth $D$, with the average physical result in blue (colored $\pm 1$ standard deviation over physical qubits) and encoded results for $n = 3, 5, 7, 9$ in different colors. As can be seen, the encoded results are better than the physical results on average, and the accuracy of the encoded results increases as $n$ increases. Subplot (b) shows the same data but with $\langle \bar{Z} \rangle$ versus $n$, with different depths $D$ shown in different colors. These curves demonstrate the exponential convergence in $n$. Note that the average physical results are shown at $n = 1$. Physical results on all qubits from which the average and standard deviation are calculated are shown in Fig.~\ref{fig:physical-qubit-results-kyiv}.
    }
    \label{fig:repetition-code-memory-experiment-results}
\end{figure*}

% =============================================================================
% \subsection{Memory experiments}
% \label{sec:memory-experiments}
% % =============================================================================
As a first experiment, we prepare the $|\bar{0}\rangle$ state of the $n$-qubit
repetition code on IBM quantum computers. We then
apply an even number $D$ of $\bar{X}$ gates and measure the expectation value $\langle
\bar{Z} \rangle$. While this sequence acts as an identity operation in an ideal, noiseless setting (where $\langle{}\bar{Z}\rangle = 1$), it allows for the investigation of gate-induced noise beyond simple decoherence. The results for this experiment on IBM Kyiv are shown in
Fig.~\ref{fig:repetition-code-memory-experiment-results}. Here, we see that the error detection results are better than the physical results on average. Further, we see that the the error detection results converge exponentially to the ideal expectation value as the code distance (number of physical qubits) increases. 

We can gain intuition for the exponential convergence by considering a model of
rate $p$ global depolarizing noise applied after each layer of the circuit, the same as in~\eqref{eqn:final-state-D-layer-depolarizing}. In this case, for any $[[n, 1]]$ code and
any logical operator $\bar{O}$, after circuit depth $D$ the error-mitigated expectation value as a function of $n$ is
\begin{equation} \label{eqn:exp-convergence-memory}
    \langle \bar{O} \rangle (n) = 
    \frac{p^D \Tr [\bar{\rho} \bar{O}]}
    {p^D + \frac{1 - p^D}{2^{n - 1}}}.
\end{equation}
Since $\Tr[\bar{\rho} \bar{O}]$ is the ideal (noiseless) expectation value, we
see the error-mitigated result converges to the ideal result exponentially in
the number of physical qubits $n$. We remark the limiting case $n = 1$
reproduces the unmitigated (noisy) expectation value $p \Tr[\rho O]$. While~\eqref{eqn:exp-convergence-memory} follows quickly from~\eqref{eqn:error-mitigated-value} and~\eqref{eqn:final-state-D-layer-depolarizing} with properties of the trace, a derivation is included in Appendix~\ref{app:derivations} for completeness.

While we see exponential convergence on hardware, the repetition code is classical --- it can only correct bit-flip errors\footnote{While we use this to motivate the study of genuine quantum codes, it is interesting to note that the chosen state and observable influence which errors are important to mitigate. Here, for example, phase-flip errors are undetectable by the bit-flip repetition code, but they do not contribute any error to the expectation value because $\Tr[ \mathcal{E}_p(\rhobar) \bar{Z} ] = \Tr[ p Z_i |\bar{0}\> \< \bar{0} | Z_i \bar{Z} + (1 - p) \rhobar \bar{Z} ]$. Since $Z_i |\bar{0}\rangle = |\bar{0}\rangle$ for any $i = 1, ..., n$ we can write this as $\Tr[ p |\bar{0}\> \< \bar{0} | \bar{Z} + (1 - p) \rhobar \bar{Z} ] = \Tr[\rhobar \bar{Z}]$.}. Furthermore, the encoding $|0\rangle \mapsto |\bar{0}\rangle$ is trivial and every stabilizer is diagonal in the $Z$-basis, so computing~\eqref{eqn:error-mitigated-value} reduces to computational basis measurements. In this sense, this presents the best-case scenario for quantum error detection. In the following experiments, we successively relax these conditions to further analyze the performance of error detection on hardware under more realistic settings.

\begin{figure}
    \centering
    \includegraphics[width=\linewidth]{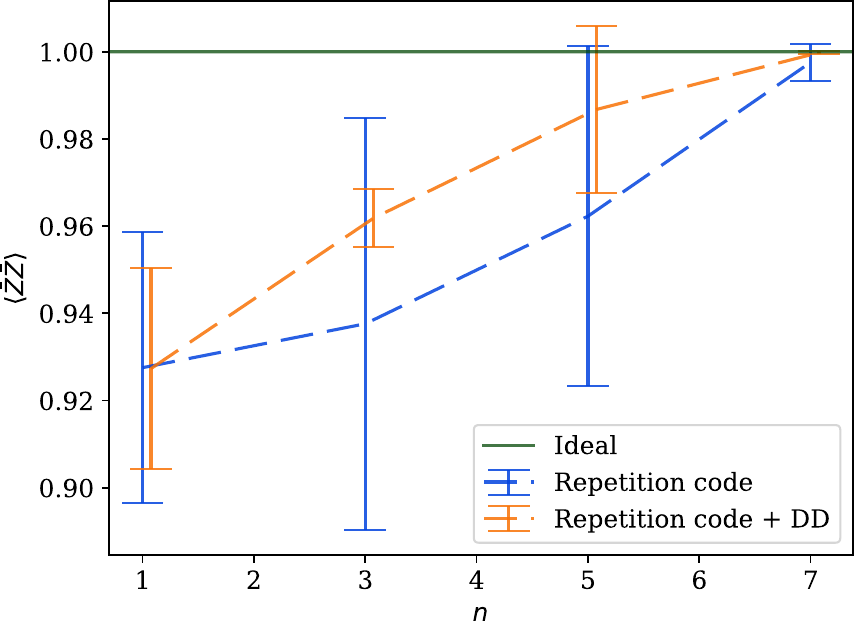}
    \caption{Logical Bell state preparation with the $n$-qubit repetition code on IBM Fez, measuring $\langle \bar{Z} \bar{Z} \rangle$. Results are shown for $n \in \{3, 5, 7\}$ with the unencoded (physical) result shown at $n = 1$. The blue curve shows the error detection results, while the orange curve shows the error detection results with $XY$-4 dynamical decoupling applied to the circuit. Error bars show one standard deviation over ten independent trials. 
    %The top plot is on \texttt{ibm\_brisbane} and the remaining plots are on \texttt{ibm\_kyiv}, all at depth $d = 0$ (i.e., executing $U (U^\dagger U)^d$ for $d = 0$ where $U$ is the logical Bell state preparation circuit). 
    %Details on qubit layouts and compiled circuit sizes are provided in Sec.~\ref{sec:methods}.
    }
    \label{fig:logical-bell}
\end{figure}

First, we consider a multi-logical-qubit computation with the repetition code, instead of a single logical qubit memory experiment. In particular, we use the $n$-qubit repetition code to prepare a logical Bell state $|\bar{\Phi}^+\rangle := (|\bar{0} \bar{0}\rangle + |\bar{1}\bar{1}\rangle) / \sqrt{2}$ and measuring $\langle \bar{Z} \bar{Z} \rangle$. The results are shown in Fig.~\ref{fig:logical-bell} for $n = 3$, $5$, and $7$, with the physical results shown at $n = 1$ for reference. As can be seen, in this experiment the error detection results are better than the physical results on average. While the results still converge to the ideal noiseless value, the rate of convergence is markedly slower than in the previous memory experiment without any two-qubit logical gates. Although the repetition code has a transversal logical CNOT, the single-qubit logical Hadamard requires a ladder of CNOTs. This makes it more probable for errors to spread. 
% (A detailed analysis of bit-flip error propagation through the logical gates of the repetition code is provided in Appendix~\ref{sec:noise-propagation}.)
Also, it increases periods in the quantum circuit where qubits are idling. To address errors accumulating during this idle windows, we also perform the experiment with dynamical decoupling (DD). As can be seen in Fig.~\ref{fig:logical-bell}, the encoded + DD results are better than just the encoded results, and the rate of convergence to the ideal value is slightly faster. 

\begin{figure*}
    \centering
    \includegraphics[width=\linewidth]{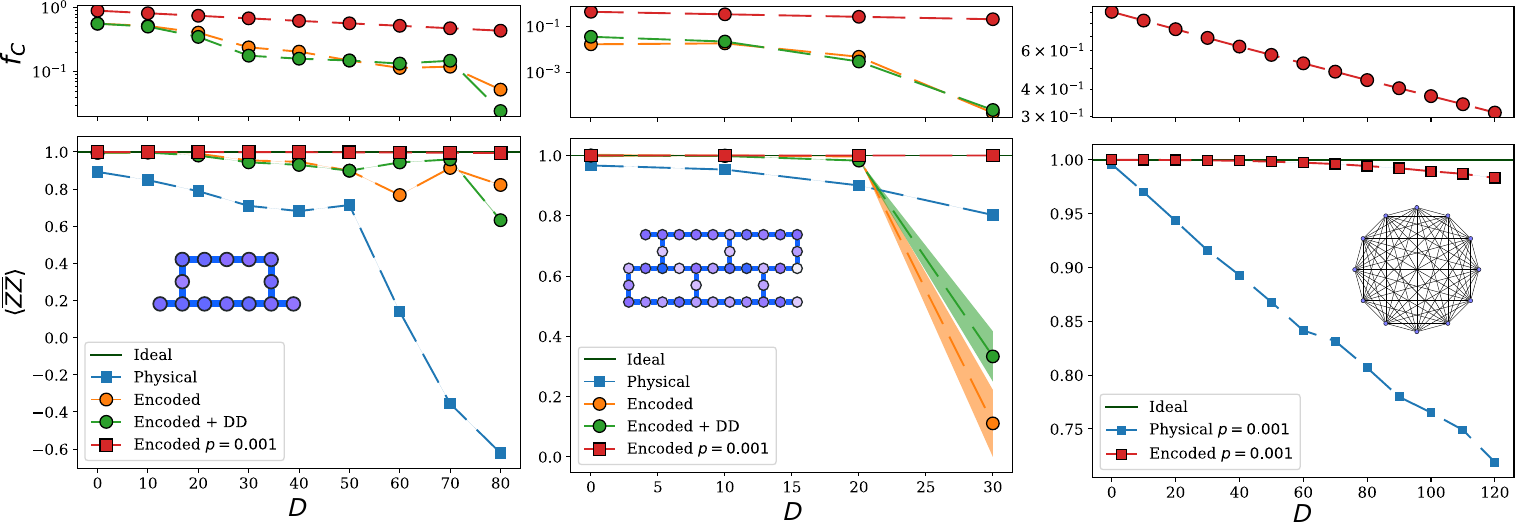}
    \caption{Results of preparing logical Bell states with the triangular color code on IBM Boston and measuring the expectation value $\langle \overline{Z} \overline{Z} \rangle$. From left to right are distances $d = 3$, $5$, and $7$, corresponding to $n = 14$, $38$, and $74$ physical qubits. For distance $d = 3$, the encoded results are better than the physical results for all executed depths $D$. For distance five, the encoded results become worse than the physical results at depth $D = 30$. Colors show one standard deviation calculated from the fraction of codewords $f_C$ sampled. For the distance seven code, no codewords were sampled at depth $D = 0$ with $10^6$ shots, and results are shown for depolarizing noise with all-to-all connectivity, also shown in the distance three and five plots for reference. As in previous experiments, circuit depths are increased by inserting an even number $D$ of $\bar{X}$ gates after the logical Bell state is prepared. Inset graphs show the device subgraphs used in the experiment (for distance seven, the graph is all-to-all on $74$ qubits, shown for $12$ nodes for visualization purposes).}
    \label{fig:tcc}
\end{figure*}

While the repetition code simplifies several experimental aspects, it is classical and only able to correct for bit-flip errors. To move beyond this, we turn to the triangular color code. Briefly, the distance $d$ triangular color code encodes $k = 1$ logical qubit using $n = (3d^2+1)/4$ physical qubits. For distance $d = 3$, the triangular color code is the seven-qubit Steane code, and higher distances generalize this by a triangular subregion of a hexagonal tiling of a two-dimensional lattice~\cite{chamberland2020triangular}. A full discussion of the triangular color code is provided in Sec.~\ref{sec:methods}. Using this code, we perform the same logical Bell state preparation experiment, this time applying $\bar{X}^D$ for even circuit depth $D$ after the logical Bell state is prepared. We emphasize the difference compared to the repetition code is a non-trivial encoding circuit mapping $|0\rangle^{\otimes n} \mapsto |\bar{0}\rangle^{\otimes k}$. The results are shown in Fig.~\ref{fig:tcc} for the distance $d = 3$, $5$, and $7$ triangular color codes on IBM Boston. At distance three ($n = 14$ physical qubits), the encoded/error detection results are better than the physical results for all circuit depths tested. At distance $5$ ($n = 38$ physical qubits) however, there is a crossover. Namely, the encoded results are better than physical for depths up to $20$, then become worse than physical after. This expected behavior illustrates what we call the \textit{pseudothreshold} --- i.e., given a particular circuit, observable, and noise model, the pseudothreshold $p^*$ is the noise level below which error detection provides superior performance and is able to exponentially suppress errors, and above which error detection provides worse performance. This is completely analogous to thresholds of quantum error correction codes, but we use the term pseudothreshold to emphasize the dependence on the particular circuit, observable, and noise model. Similar to the previous experiment, we also apply dynamical decoupling to the logical circuits. While the performance with dynamical decoupling is better at some depths, it is less emphatic and less clear than the results we obtained for the repetition code. Finally, we also implemented the distance $7$ triangular color code corresponding to $74$ physical qubits, but did not sample any codewords on hardware after $10^6$ shots. To study the code at this distance, we implemented a noise model of single-qubit depolarizing noise after every moment (parallel layer of circuit operations), for a noise rate comparable to hardware. In this case, we consider all-to-all connectivity, instead of the heavy hex connectivity provided by IBM Boston. The noisy simulation results indicate that error detection outperforms the physical experiment at all tested circuit depths. The same noisy simulations were performed for the distance $3$ and distance $5$ experiments, and are shown for comparison on the respective subplots in Fig.~\ref{fig:tcc}. 

\begin{figure}
    \centering
    \includegraphics[width=\linewidth]{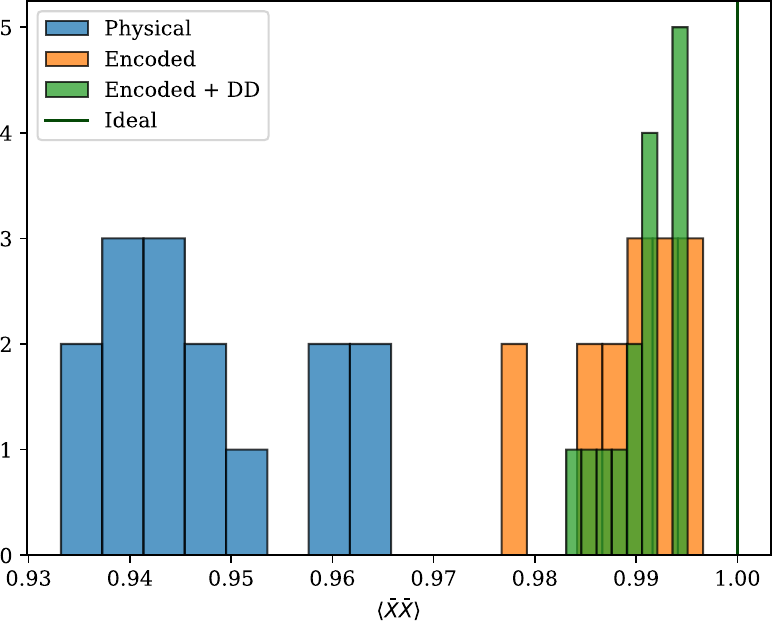}
    \caption{Histogram of results for preparing the logical Bell state $|\bar{\Phi}^{+}\rangle$ on IBM Fez and measuring $\langle \bar{X} \bar{X} \rangle$ with the distance $d = 3$ triangular color code (Steane code).}
    \label{fig:bell-steane-xx}
\end{figure}

Although we chose $\overline{ZZ}$ for the observable in this experiment, we could have equally well chosen $\overline{XX}$ or $\overline{YY}$ since they are related by a transversal basis rotation in the triangular color code. To illustrate this, Fig.~\ref{fig:bell-steane-xx} shows the distance three experiment performed with the $\overline{XX}$ observable. The only difference in the circuit is a transversal Hadamard applied before the computational basis measurement. A histogram of results over $15$ independent experiments on IBM Fez show that the error detection performance is better than physical on average, just as with the $\overline{ZZ}$ observable used for the previous experiment.

\begin{figure}
    \centering
    \includegraphics[width=\linewidth]{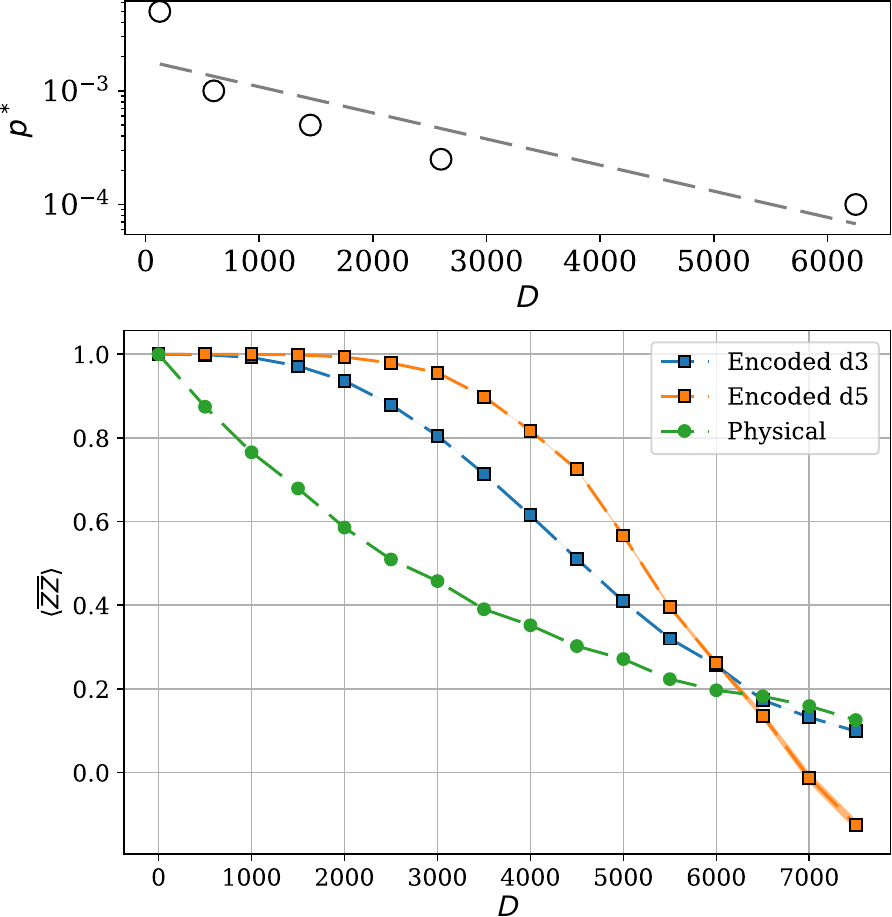}
    \caption{Estimated pseudothreshold $p^*$ (top panel) for the triangular color code Bell state preparation experiment, applying $\bar{X}^D$ for a given depth $D$. The pseudothreshold $p^*$ is defined as the the physical noise rate $p$ below which error detection outperforms the physical experiment. Markers show calculated thresholds and the dashed line shows an exponential fit. Thresholds are calculated by simulating the experiment with different code distances and estimating the crossover point (bottom plot), the same method for estimating thresholds of quantum error correction codes.}
    \label{fig:pseudo-threshold}
\end{figure}

Finally, we explore the pseudothreshold behavior exhibited in Fig.~\ref{fig:tcc} more deeply. Again, we define the pseudothreshold as the physical noise rate $p^*$ below which error detection provides better performance, provided a given circuit, observable, and noise model. Just as in error correction, (pseudo)thresholds are difficult to evaluate analytically, and this can only be done in special cases. To explore the pseudothreshold here, we consider the same logical Bell state preparation experiment with the triangular color code and a model of rate $p$ single-qubit depolarizing noise applied after every parallel layer of operations. To estimate the pseudothreshold, we simulate the physical experiment and the encoded experiment at multiple distances, then estimate the crossover point. The results are shown in Fig.~\ref{fig:pseudo-threshold}. As expected, we see that the pseudothreshold $p^*$ drops exponentially in the depth of the circuit. This sub-threshold region characterizes the circuit volumes and noise rates for which error detection provides better-than-physical performance.

% =============================================================================
\section{Methods} \label{sec:methods}
% =============================================================================

\begin{figure}
    \centering
    \includegraphics[width=\linewidth]{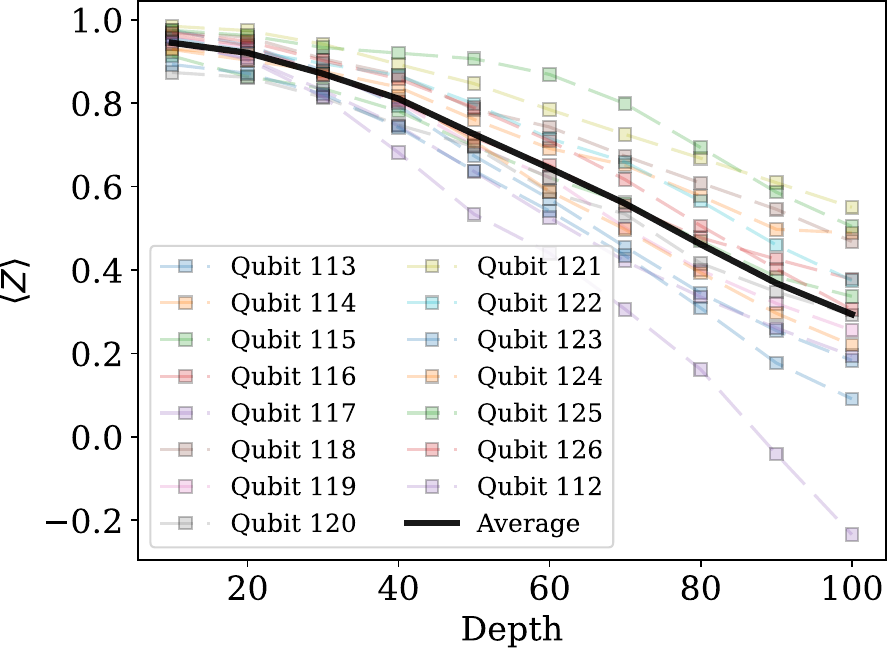}
    \caption{Individual physical qubit results for the repetition code memory
    experiment on IBM Kyiv (Fig.~\ref{fig:repetition-code-memory-experiment-results}).}
    \label{fig:physical-qubit-results-kyiv}
\end{figure}

Unless otherwise mentioned, we performed physical experiments with $10^4$ shots and logical experiments with $10^5$ shots on hardware. All hardware experiments were performed with IBM quantum computers. For each experiment, we select the best subset of qubits to run on from the latest calibration data. Qubits generally exhibit non-trivial variance over space and time --- Fig.~\ref{fig:physical-qubit-results-kyiv} shows the individual qubit results from the average physical results calculated in Fig.~\ref{fig:repetition-code-memory-experiment-results}. For experiments, we choose contiguous qubit subsets to minimize the number of two-qubit gates required in circuits. For the repetition code these subsets are simply lines of qubits for both memory and logical Bell state preparation experiments. For the triangular color code, we choose adjacent hexagons of qubits on the device. Noisy simulations were performed with a Clifford circuit simulator~\cite{Gidney_2021} assuming all-to-all connectivity and a model of single-qubit depolarizing noise applied after every moment.

\begin{figure}
    \centering
    \includegraphics[width=\linewidth]{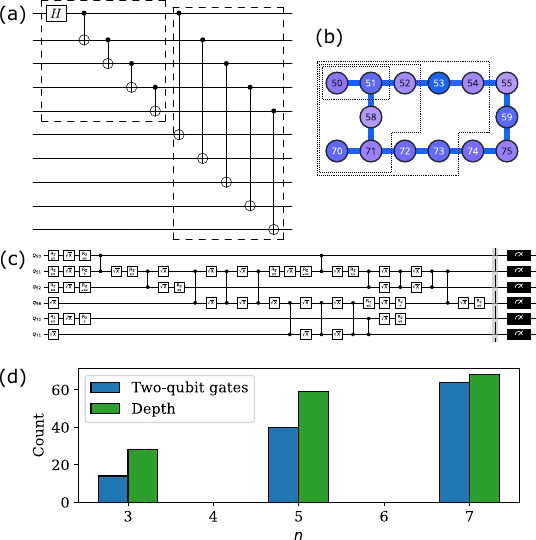}
    \caption{Methods for the repetition code Bell state experiment (Fig.~\ref{fig:logical-bell}). (a) Circuit used to prepare the logical Bell state from the physical zero state (which is also the logical zero state for the repetition code). The first dashed box implements the mapping $|\bar{0}\rangle \mapsto (|\bar{0}\rangle + |\bar{1}\rangle) / 2$ for the $n$-qubit repetition code, and the second dashed box shows the logical CNOT, shown for $n = 5$. (b) Example selection of qubits on the IBM heavy hex topology, shown here for IBM Fez. Qubit colors show $T_2$ times (lighter is better) and edge connections show two-qubit error rates (darker is better). Dashed boxes show qubit selections for the physical experiment and encoded experiment with $n = 3$, $5$, and $7$ qubits. (This $n$ is the number of physical qubits per logical qubit; with two logical qubits, the number of physical qubits used is twice this number.) (c) An example compilation of the circuit (a) to the heavy hex topology (b), with SWAP gates are introduced for qubit routing. (d) Circuit statistics showing the number of two-qubit gates and circuit depths for $n = 3$, $5$, and $7$. The largest $n = 7$ circuit has $2n = 14$ total physical qubits, $64$ two-qubit (CZ) gates, and a depth of $68$. Note that compilation can be random, and different qubit subsets were chosen at different times based on device calibration, but these numbers are representative for all circuits used in the experiment.}
    \label{fig:repetition-bell-methods}
\end{figure}

% \textbf{Repetition code.}
The $n$-qubit repetition code is an $[[n, 1, n]]$ code (for odd $n$) with
stabilizer generators $\mathcal{S} = \langle Z_1 Z_2, Z_2 Z_3, \dots,
Z_{n-1} Z_n \rangle$ and logical operators $\bar{Z} = Z_1$ and $\bar{X} =
X_1 X_2 \cdots X_n$~\cite{girvin2023introduction, albert2023repetition}. As
a classical code, it can only detect and correct bit-flip ($X$) errors. The
logical CNOT gate is transversal $\overline{\text{CNOT}} = \prod_i
\text{CNOT}_{i, i+n}$. To implement the mapping $|\bar{0}\rangle \mapsto (|\bar{0}\rangle + |\bar{1}\rangle) / 2$ for the logical Hadamard, we use a physical Hadamard and sequence of CNOT gates. See Fig.~\ref{fig:repetition-bell-methods}(a). This gate is not fault-tolerant and can propagate bit-flip errors. While the Bell state preparation circuit can be embedded directly into a two-dimensional ``ladder'' topology, embedding into the heavy hex topology requires overhead in the form of SWAP gates. An example qubit subset is shown in Fig.~\ref{fig:repetition-bell-methods}(b) and corresponding compilation in Fig.~\ref{fig:repetition-bell-methods}, which illustrates this overhead. Figure~\ref{fig:repetition-bell-methods}(d) shows circuit statistics for all $n = 3$, $5$, and $7$ used in the experiment.

% \textbf{Triangular color code.}

\begin{figure}
    \centering
    \includegraphics[width=\linewidth]{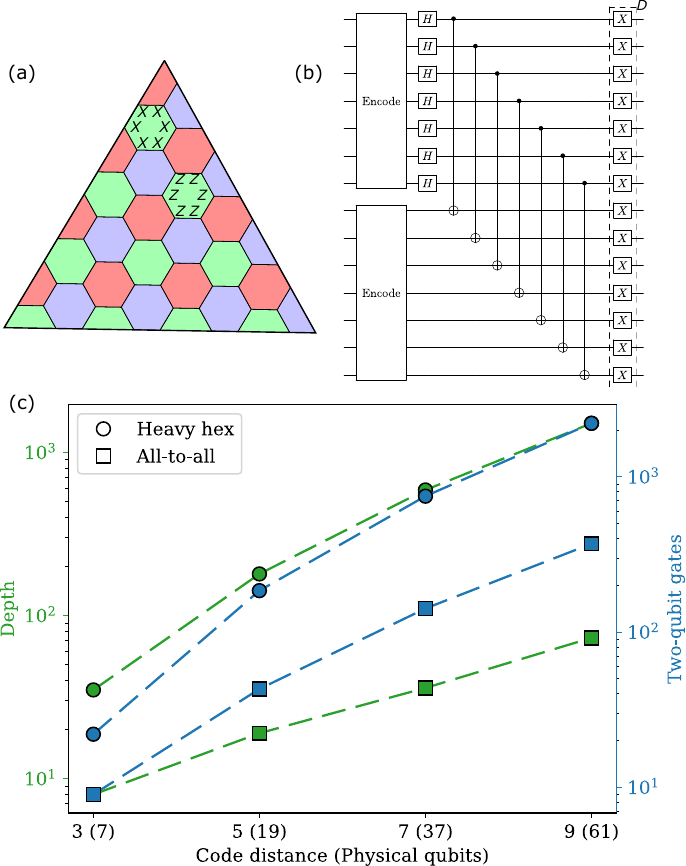}
    \caption{Methods for the triangular color code Bell state experiment (Fig.~\ref{fig:tcc}). (a) The triangular color code is defined by a hexagonal tiling of a triangular lattice. Qubits live on vertices of the lattice. As a CSS code, each face contributes an $X$ type stabilizer and a $Z$ type stabilizer. (b) High level circuit diagram for the experiment. First, the logical $|\bar{0} \bar{0}\rangle$ state is prepared by a unitary encoding circuit, then a logical Bell state is prepared by logical Hadamard and logical CNOT, which are both transversal for the triangular color code. Finally, a layer of $\bar{X}$ gates is applied to each logical qubit for a depth $D$. (c) Number of two-qubit gates and circuit depths for unitarily preparing the logical zero state of the triangular color code from the all zero physical state. Note that this corresponds to one ``Encode'' gate in (b). Subsets of the heavy hex topology are shown in the inset of Fig.~\ref{fig:tcc}.}
    \label{fig:tcc_encodings}
\end{figure}

The distance $d$ triangular color code is an $[[n, 1, d]]$ code with $n =
(3d^2 + 1)/4$ physical qubits arranged on a triangular lattice with
three-colorable faces~\cite{bombin2013introduction, chamberland2020triangular,
albert2024honeycomb}. See Fig.~\ref{fig:tcc_encodings}(a). Unlike the repetition code, it is a quantum code capable
of correcting both bit-flip and phase-flip errors on any physical qubit. For distance $d = 3$ (the
Steane code, $n = 7$), the stabilizer group is generated by six operators (three
$X$-type and three $Z$-type) corresponding to the faces of the lattice, and this generalizes to higher distances with larger lattices. In the triangular color code, both
the logical Hadamard and logical CNOT are
transversal. The high-level circuit diagram for the Bell state experiment is shown in Fig.~\ref{fig:tcc_encodings}(b). Unlike the repetition code in which logical state preparation $|0\rangle^{\otimes n} \mapsto |\bar{0}\rangle^{\otimes k}$ is trivial, the triangular color code requires an encoding method. Generally this can always be done for error correcting codes by measuring each stabilizer and applying an appropriate correction to ensure the $+1$ eigenstate of all stabilizers --- i.e., the $|\bar{0}\rangle$ state --- is prepared. This is equivalent to a round of error correction applied to the $|0\rangle^{\otimes n}$ physical state. On IBM quantum computers we are currently unable to reliably perform this encoding technique, so we instead implement a unitary circuit $U$ such that $U|0\rangle^{\otimes n} = |\bar{0}\rangle^{\otimes k}$. This can always be done by performing Gaussian elimination on the stabilizer tableau (check matrix) of the code and implementing the corresponding circuit operations. Fig.~\ref{fig:tcc_encodings}(c) shows the number of two-qubit gates and circuit depths for unitarily encoding the logical zero state of the distance $d$ triangular color code, both on all-to-all connectivity used in noisy simulations and on the heavy hex topology on IBM quantum computers. This plot shows the relative complexity of unitary encoding as well as the overhead incurred by embedding to the heavy hex architecture.

% =============================================================================
\section{Conclusion}
\label{sec:conclusion}
% =============================================================================

In this work, we have experimentally demonstrated that quantum error detection can improve the
accuracy of expectation values exponentially in the number of physical qubits,
provided the hardware noise rate is below a threshold. This mirrors the
threshold behavior observed in quantum error correction and provides a key opportunity for performing quantum error detection experiments on hardware. For the repetition code memory experiments, we observe clear exponential
convergence as $n$ increases, consistent with the simple theoretical
prediction of~\eqref{eqn:exp-convergence-memory}. While the Bell state preparation
experiments also show the opportunities for quantum error detection, they exhibit the challenges of scaling as well. In particular, as the code size grows,
 compiled circuits become deeper and introduce additional noise that can push
the effective error rate above the pseudothreshold.

We expect that optimized encoding circuits would further improve our experimental results. Further, we expect that non-unitary encoding circuits would improve our results when experimentally available, although these also involve additional noise arising from noisy ancilla state preparation and syndrome measurements. It is an interesting future direction to develop quantum error correction codes tailored to particular hardware to alleviate these challenges. While we characterize the post-selection rate (fraction of codewords $f_C$ measured), we did not normalize the improvement of quantum error detection by the additional resources (shots) used.

Our benchmarks show that quantum error detection holds strong promise as an unbiased error mitigation technique. If new techniques are introduced to alleviate the exponential sampling
overhead and the exponential classical processing, even larger-scale, application-oriented experiments with quantum error detection will be able to be performed. Such techniques have been proposed in recent literature, e.g. generalized
subspace expansion~\cite{yoshioka2022generalized} and logical shadow
tomography~\cite{hu2022logical}, and are of significant interest for future work.

\vspace{1.0em}

% =============================================================================
\section*{Code and data availability} 
% =============================================================================

Code to reproduce experiments is available at~\cite{LaRose_2026}.
\vspace{0.5em}

% =============================================================================
\section*{Acknowledgments} 
% =============================================================================

This work was supported by Wellcome Leap as part of the Q4Bio Program. YLF acknowledge support by the project PID2023-149969NA-100 (SEQUOIA) funded by the Spanish Agencia Estatal de Investigación MICIU/AEI/10.13039/501100011033, and by a 2025
Leonardo Grant for Scientific Research and Cultural Creation from the BBVA Foundation.
We thank IBM for providing access to their quantum computers through the Quantum Research Credits Award program. The views expressed in this article are those of the authors and do not necessarily reflect those of IBM.

% \vspace{0.5em}

% % =============================================================================
% \textbf{Author contributions} 
% % =============================================================================

% \vspace{0.5em}

% % =============================================================================
% \textbf{Competing interests} The authors declare no competing interests.
% =============================================================================

\bibliography{refs.bib}

% =============================================================================
\appendix
% =============================================================================

\section{Derivations} \label{app:derivations}

To see why~\eqref{eqn:exp-convergence-memory} is true, the error-mitigated value
is defined as
\begin{equation} \label{eqn:mitigated-expectation-value-function-of-n}
    \langle \bar{O} \rangle (n) := 
    \frac{\Tr[\Pi \mathcal{E}_p (\bar{\rho}) \Pi^\dagger \bar{O}]}
    {\Tr[\Pi \mathcal{E}_p (\bar{\rho}) \Pi^\dagger]},
\end{equation}
where $\Pi :=
\prod_{S \in \mathcal{S}} (I + S) / 2$ is the projector onto the codespace, $\mathcal{S}$ is a generating set for the stabilizer group, and
$\mathcal{E}_p := p^D I / 2^n + (1 - p^D) \bar{\rho}$ is the action of rate $p$
depolarizing noise on the encoded state $\bar{\rho}$ after $D$ circuit layers. Using the facts that $[\Pi, \bar{O}] = 0$, $\Pi =
\Pi^\dagger$, and $\Pi^2 = \Pi$, the numerator
of~\eqref{eqn:mitigated-expectation-value-function-of-n} is
\begin{align}
    \Tr [ \Pi \mathcal{E}_p (\bar{\rho}) \Pi^\dagger \bar{O}] &= \frac{1 - p^D}{2^n} \Tr[ \Pi \bar{O}] + p^D \Tr [ \bar{\rho} \Pi \bar{O}] = p^D \Tr [ \bar{\rho} \bar{O}] .
\end{align}
The second equality follows because every term in $\Pi \bar{O}$ is a Pauli so that $\Tr[ \Pi \bar{O} ] = 0$, and because $\bar{\rho}$ is a codeword so that $\bar{\rho} \Pi = \bar{\rho}$. Similarly, the denominator of~\eqref{eqn:mitigated-expectation-value-function-of-n} is
\begin{equation}
    (1 - p^D) \Tr [\Pi ] / 2^n + p^D \Tr [ \bar{\rho} \Pi] = (1 - p^D) / 2^{n - 1} + p^D .
\end{equation}
This follows because $\Pi = \frac{1}{2^{n-1}} \sum_{S \in \mathcal{S}} S$ and every non-identity Pauli is traceless, so $\Tr[\Pi] = \frac{1}{2^{n-1}} \Tr[I] = \frac{2^n}{2^{n-1}} = 2$, giving $\Tr[\Pi]/2^n = 1/2^{n-1}$. Similarly, $\Tr[\bar{\rho} \Pi] = 1$ since $\bar{\rho}$ is a codeword. Combining the numerator and denominator yields~\eqref{eqn:exp-convergence-memory}.

\end{document}